\documentclass[conference]{IEEEtran}
\IEEEoverridecommandlockouts
\usepackage{cite}
\usepackage{amsmath,amssymb,amsfonts}
\usepackage{algorithmic}
\usepackage{graphicx}
\usepackage{textcomp}
\usepackage{xcolor}
\def\BibTeX{{\rm B\kern-.05em{\sc i\kern-.025em b}\kern-.08em
    T\kern-.1667em\lower.7ex\hbox{E}\kern-.125emX}}
				
\usepackage{upgreek} 	
\usepackage{amssymb}
\usepackage{amsthm} 

\newtheorem{Corollary}{Cor.}
\newtheorem{Definition}{Def.}
\newtheorem{Example}{Ex.}
\newtheorem{Proposition}{Prop.}
\newtheorem{Remark}{Rem.}
\newtheorem{Theorem}{Thm.}

\newcommand{\gedit}[1]{\textcolor{black}{#1}}

\begin{document}

\title{On Large-Scale Graph Generation with Validation of Diverse Triangle Statistics at Edges and Vertices
}

\author{\IEEEauthorblockN{Geoffrey Sanders, Roger Pearce, Timothy La Fond \thanks{This work was performed under the auspices of the U.S. Department of Energy by Lawrence Livermore National Laboratory under Contract DE-AC52-07NA27344 and was supported by the LLNL-LDRD Program under Project No. 17-ERD-024, LLNL-CONF-748352}}
\IEEEauthorblockA{\textit{Center for Applied Scientific Computing (CASC)} \\
\textit{Lawrence Livermore National Laboratory}\\ 
Livermore, CA, USA \\
sanders29@llnl.gov, pearce7@llnl.gov, lafond1@llnl.gov}
\and
\IEEEauthorblockN{Jeremy Kepner}
\IEEEauthorblockA{\textit{ Lincoln Laboratory Supercomputing Center} \\
\textit{MIT Lincoln Laboratory}\\
Lexington, MA, USA \\
kepner@ll.mit.edu}
}

\maketitle


\begin{abstract}

Researchers developing implementations of distributed graph analytic algorithms require graph generators that yield graphs sharing the challenging characteristics of real-world graphs (small-world, scale-free, heavy-tailed degree distribution) with efficiently calculable ground-truth solutions to the desired output.  
Reproducibility for current generators \cite{graph500} used in benchmarking are somewhat lacking in this respect due to their randomness: the output of a desired graph analytic can only be compared to expected values and not exact ground truth.
Nonstochastic Kronecker product graphs \cite{Weischel1962} meet these design criteria for several graph analytics.
Here we show that many flavors of triangle participation can be cheaply calculated while generating a Kronecker product graph.
 
Given two medium-sized scale-free graphs with adjacency matrices $A$ and $B$, their Kronecker product graph has adjacency matrix $C = A \otimes B$.
Such graphs are highly compressible: $|{\cal E}|$ edges are represented in ${\cal O}(|{\cal E}|^{1/2})$ memory and can be built in a distributed setting from small data structures, making them easy to share in compressed form. 
Many interesting graph calculations have worst-case complexity bounds  ${\cal O}(|{\cal E}|^p)$ and often these are reduced to ${\cal O}(|{\cal E}|^{p/2})$ for Kronecker product graphs, when a Kronecker formula can be derived yielding the sought calculation on $C$ in terms of related calculations on $A$ and $B$.  

We focus on deriving formulas for {\em triangle participation at vertices}, ${\bf t}_C$, a vector storing the number of triangles that every vertex is involved in, and {\em triangle participation at edges},  $\Delta_C$, a sparse matrix storing the number of triangles at every edge.   
When factors $A$ and $B$ are undirected, $C$ is also undirected.   
In the case when both factors have no self loops we show ${\bf t}_C = 2 {\bf t}_A \otimes {\bf t}_B$, $\Delta_C = \Delta_A \otimes \Delta_B$.   Moreover, we derive the respective formulas when $A$ and $B$ have self loops, which boosts the triangle counts for the associated vertices/edges in $C$.  We additionally demonstrate strong assumptions on $B$ that allow the truss decomposition of $C$ to be derived cheaply from the truss decomposition of $A$. 

We extend these results and show Kronecker formulas for triangle participation in both directed graphs and undirected, vertex-labeled graphs.   In these classes of graphs each vertex / edge can participate in many different types of triangles.   

\end{abstract}

\section{Introduction}
In recent work \cite{Kepner2018GABB}, extremely large synthetic power-law Kronecker graphs \cite{Weischel1962} are generated in an essentially communication-free implementation for the primary purpose of validating graph calculation implementations on benchmarks where the answer is known exactly.   
The (non-stochastic) Kronecker approach leverages ideas an observations from previous work on stochastic Kronecker graph generators \cite{Chakrabarti2004, Leskovec2005, kepner2011graph}, with the added benefit that the ground truth of many local and global graph statistics are efficiently calculated during the generation process. In \cite{Leskovec2010KronFit}, several properties like graph diameter are discussed for both the non-stochastic and stochastic cases.

\begin{figure}[t]
   \centering
   \includegraphics[width=3.6in]{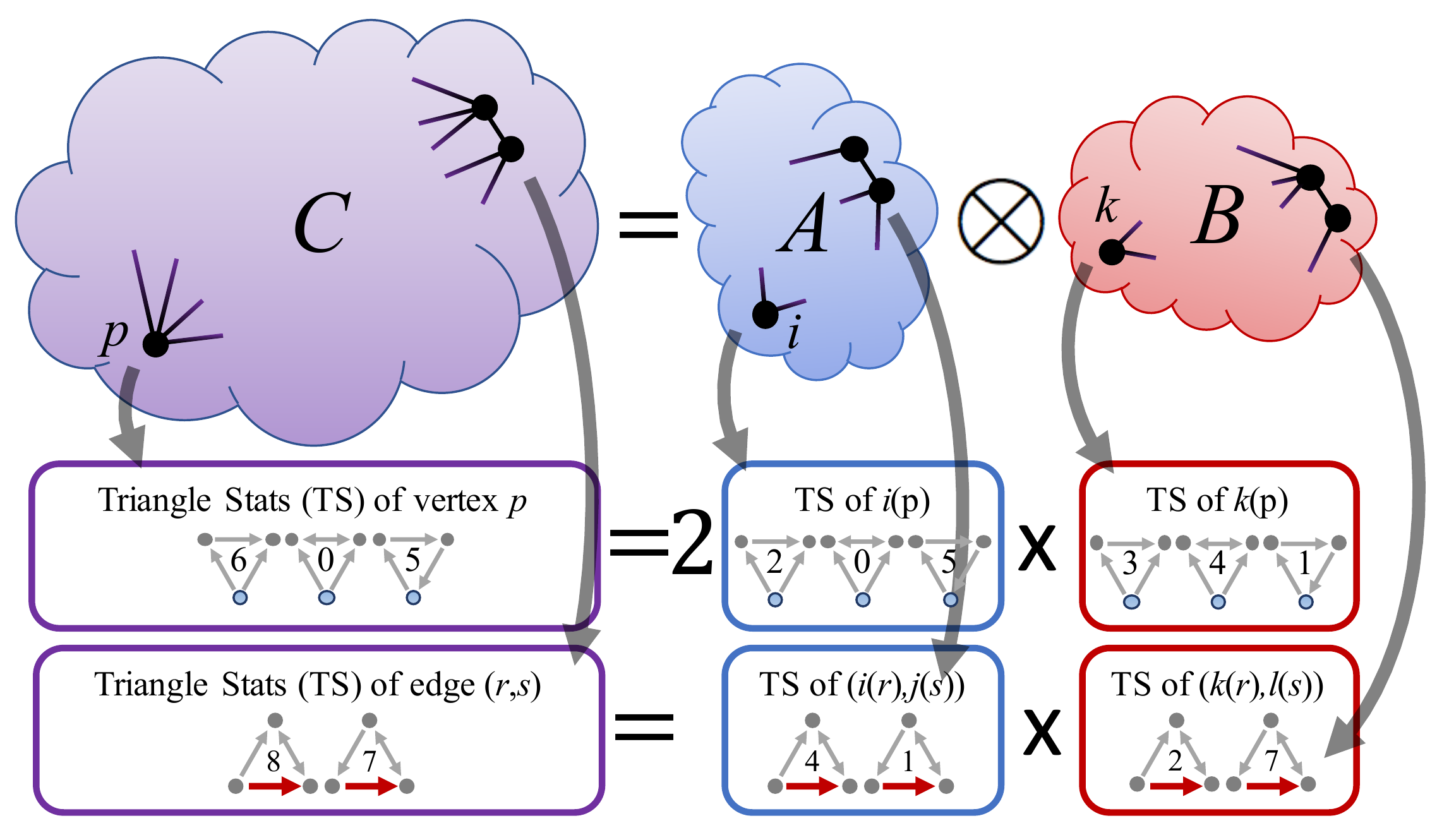} 
   \caption{Under mild assumptions on $A$ and $B$ (e.g. $A$, $B$ have no self loops, $B$ is undirected) diverse triangle statistics of $C = A \otimes B$ are simply a constant times Kronecker products of the triangle statistic vectors / matrices associated with $A$ and $B$, allowing for efficient exact local triangle statistics to be calculated during graph generation.   }
   \label{fig:hadamard}
\end{figure}

We form a graph ${\cal G}_C$ whose adjacency matrix is a Kronecker product \cite{VanLoan2000,Weischel1962, HornAndJohnson} of two much smaller factors, $C = A \otimes B =$
$$
\left(
\begin{array}{cccc}
A_{11} B & A_{12} B & \cdots  & A_{1,n_A} B \\
A_{21} B & A_{22} B & \cdots  & A_{2,n_A} B \\
\vdots & \vdots & \ddots & \vdots \\
A_{n_A,1} B & A_{n_A,2} B & \cdots & A_{n_A,n_A} B
\end{array}
\right),
$$
as this framework provides the ability to calculate many (normally expensive) graph statistics for $C$ cheaply from associated statistics on $A$ and $B$.   
A polynomial time graph calculation has the potential to be done at a square-root of the \gedit{general worst-case} cost and inline with the graph generation process.   

Consider triangle counting as an example.   Suppose the number of edges in $A$ and $B$, $|{\cal E}_A|$ and $|{\cal E}_B|$, are both ${\cal O}(10^6)$ then the number of edges in $|{\cal E}_C|$ is ${\cal O}(10^{12})$. 
We want to compute  the number of triangles in $C$, $\tau(C)$. 
\gedit{For a general graph of this size, computing has $\tau(C)$ worst-case complexity ${\cal O}(|{\cal E}_C|^{3/2})$ \cite{Chiba1985} or ${\cal O}(10^{18})$ (and $\tau(C)$ itself could be ${\cal O}(10^{18})$).}
However, the worst-case bound for counting triangles in a $C$ of this form, using the Kronecker formula $6\tau(A) \tau(B)$, is ${\cal O}(|{\cal E}_C|^{3/4})$ or ${\cal O}(10^{9})$; the number of triangles in a trillion-edge graph is computed sublinearly -- in the worst case.   If $A$ and $B$ are sparse, real-world, power-law graphs, the actual complexity of an implementation leveraging heuristics can be as low as ${\cal O}(\tau(A) + \tau(B))$, which is often significantly lower than ${\cal O}(10^{9})$ \cite{samsi2017static, pearce2017triangle},
\gedit{even in cases where $\tau(C)$ is as high as possible for a graph of its size, ${\cal O}(10^{18})$.}

This class of generators allows researchers to validate implementations of graph calculations that ignore the Kronecker framework on problems where the solution is known, and gain confidence in the implementations' application to extremely large real-world graphs, where the solution is fundamentally unknown and the only hope of validation is the agreement between two or more competing implementations.
\gedit{
\begin{Remark}
The stochastic Kronecker graphs from \cite{Chakrabarti2004, graph500} are demonstrated in \cite{Leskovec2010KronFit, Pinar2011SKG} to have relatively few triangles compared to large sparse real-world graphs.   This is due to the independence of edges in the stochastic model and their extremely low combined probability for most vertex triplets.   The non-stochastic Kronecker graphs we consider here are fundamentally different: they do not necessarily suffer from unreasonably low triangle counts and we give many cases throughout this paper where non-stochastic Kronecker graphs have a high number of triangles.   Additionally, our formulas allow tuning of local triangle counts by adding/deleting traingles and self-loops from the input factors.
\end{Remark}
}

Triangle-related graph analysis is extremely important to many applications. 
In undirected graphs, triangle participation at vertices (or {\em triangle degree}) is a common expensive graph statistic for metrics like the local clustering coefficient of a vertex \cite{WattsStrogatz1998}.   
Similarly, triangle participation at edges is used for clustering coefficient of an edge.   
Both types of participation are additionally used in several less-local graph analytics like improved clustering \cite{Berry2011ToleratingTC}, truss decompositions \cite{Cohen2008ktruss, voegele2017parallel, bisson2017static, smith2017truss, date2017collaborative}, and realistic graph generation.
Furthermore, in directed graphs or labeled graphs, diverse triangle statistics that count various types of triangles are calculable at every vertex and edge (see Figs~\ref{fig:vtris}, \ref{fig:etris}, and \ref{fig:ltris}) and these statistics are useful for several interesting applications like motif-based clustering \cite{Benson163} or  pattern detection \cite{Reza2017}.
All forms of triangle participation are likely attractive topological features in supervised/unsupervised machine learning applications \cite{Henderson2012RolX, Grover2016node2vec}.

There is currently significant research effort towards developing algorithms and systems capable of computing large-scale triangle counting, participation, and enumeration.   These efforts include implementations in MapReduce \cite{cohen2009, burkhardt2016}, leveraging GPUs \cite{wang2016, bisson2017static, date2017collaborative}, in shared memory \cite{tom2017exploring}, utilizing linear algebraic kernels \cite{gilbert2015, wolf2017fast, hutchison2017distributed, low2017first}, and several other graph HPC implementations \cite{voegele2017parallel, pearce2017triangle, hu2017trix, mowlaei2017triangle}.
A recent workshop, IEEE HPEC 2017 Graph Challenge \cite{bisson2017static}, was organized to accelerate the progress of these efforts via cross-collaboration.

In this paper, we derive several new Kronecker formulas for diverse triangle participation counts of all vertices and edges in several classes of graphs.   Our contributions include: 
\begin{itemize}
\item[(a)] formulas for triangle participation of all edges and vertices in an undirected Kronecker product graph, in cases where both factors have no self loops, where only one factor has self loops, or where both factors have self loops.
\item[(b)] formulas for participation of all edges and vertices in the many types of directed triangles in a directed Kronecker product graph, in cases where one factor is directed (nonsymmetric) without self loops and the other factor is undirected (symmetric) and possibly contains self loops.
\item[(c)] formulas for participation of all edges and vertices in the many types of vertex-labeled triangles in an undirected vertex-labeled Kronecker product graph, in cases where one factor is vertex-labeled without self loops and the other factor is unlabeled and possibly contains self loops.
\item[(d)] Several implications of these formulas regarding properties of degree and triangle distributions for  Kronecker product graphs.
\item[(e)] A strategy for generating graphs with known truss decomposition.
\item[(f)] Several simple examples for validating and checking these formulas.
\end{itemize} 


\section{Preliminaries}

Matrices formed by Kronecker products are block structured and we define some convenience functions to write the index maps compactly.  For a block-structured array with block-size $n$, we define functions that, for a given {\em global index} $i$,  retrieve the {\em block number}, $\alpha_n(i)$, and the {\em intra-block index} $\beta_n(i)$.
\begin{eqnarray*}
\label{eqn:alphabeta}
\alpha_n (i) & = & \left\lfloor (i-1) / n \right\rfloor +1, \\
\beta_n(i) & = &  \left[(i-1) \% n \right] +1.
\end{eqnarray*}
The inverse of $i \rightarrow (\alpha_n(i), \beta_n(i))$ is 
$$
\label{eqn:gamma}
\gamma_n(x,y) = (x-1) n + y,
$$
in the sense that 
$
i = \gamma_n(\alpha_n(i), \beta_n(i)).
$

\begin{Definition}{\bf (Kronecker Product \cite{VanLoan2000, Weischel1962, HornAndJohnson})}
\label{def:kron}
Let $A \in \mathbb{R}^{m_A \times n_A}$ and $B \in \mathbb{R}^{m_B \times n_B}$.   The {\em Kronecker Product} of $A$ and $B$ is $(A \otimes B) \in \mathbb{R}^{(m_A m_B) \times (n_A n_B) }$ and has entries 
$$
(A \otimes B)_{pq} = 
\left(A_{\alpha_{m_B}(p), \alpha_{n_B}(q)}\right)\left( B_{\beta_{m_B}(p) ,  \beta_{n_B}(q)} \right)
$$
for $1 \leq p \leq (m_A m_B)$ and $1 \leq q \leq (n_A n_B)$, or, equivalently,
$$
(A \otimes B)_{\gamma_{m_B}(i,k) , \gamma_{n_B}(j,l)} = A_{ij} B_{kl},
$$
for $1\leq i \leq m_A, 1 \leq j \leq n_A, 1\leq k \leq m_B,$ and $1 \leq l \leq n_B$.
\end{Definition}


\begin{Proposition}{\bf (Properties of the Kronecker Product \cite{VanLoan2000, Weischel1962, HornAndJohnson})}
\label{prop:kformulas}
\vspace{-.2cm}
\begin{itemize}
\item [(a)] {\sc Scalar Multiplication.}
For any $a_1, a_2 \in \mathbb{R}$,   
$$
(a_1 a_2) (A_1 \otimes A_2) = (a_1 A_1) \otimes (a_2 A_2).
$$
\item[(b)] {\sc Distributivity.}
\begin{eqnarray*}
(A_1 + A_2) \otimes A_3 & = &(A_1 \otimes A_3)+ (A_2 \otimes A_3)
\qquad \mbox{and} \\ 
A_1 \otimes (A_2 + A_3) & = & (A_1 \otimes A_2)+ (A_1 \otimes A_3).
\end{eqnarray*}
\item[(c)] {\sc Tranposition.}
$
(A_1 \otimes A_2)^t = (A_1^t \otimes A_2^t).
$
\item[(d)] {\sc Matrix-Matrix Multiplication.}
When $n_{A_1} = m_{A_3}$ and $n_{A_2} = m_{A_4}$,
$$
(A_1 \otimes A_2) (A_3 \otimes A_4) = (A_1 A_3) \otimes (A_2 A_4).
$$
\end{itemize}

\end{Proposition}

\begin{Definition}{\bf (Haddamard Product \cite{HornAndJohnson})}
\label{def:hadd}
Let $A, B \in \mathbb{R}^{m \times n}$.   The {\em Haddamard Product} of $A$ and $B$ is $(A \circ B) \in \mathbb{R}^{m \times n }$, with
$$
(A \circ B)_{ij} = A_{ij} B_{ij}
$$
for $1 \leq i \leq m$ and $1 \leq j \leq n$. 
\end{Definition}

\begin{Definition}{\bf (Standard Matrix and Vector Objects)}
Given $A \in \mathbb{R}^{n_A \times n_A}$, $O_A$ is the {\em matrix of all zeros} and  $I_A$ is the {\em identity matrix}, both with the same size of $A$. 
Constant vectors ${\bf 0}_A, {\bf 1}_A \in \mathbb{R}^{n_A}$, are the {\em vector of all zeros}, and the {\em vector of all ones}, both of dimension $A$.  
\end{Definition}
We define some diagonal operators of square matrices in terms of the Haddamard product and recall several useful formulas regarding Haddamard products, as they simplifiy many derivations in the rest of the paper.
\begin{Definition}{\bf (Matrix Diagonal Operators)}
Given $A \in \mathbb{R}^{n_A \times n_A}$,  the matrix $D_A = I_A\circ A$ is the {\em diagonal entries} of $A$.    The {\em diagonal operator} is $\mbox{diag}(A) := (I_A \circ A) {\bf 1}_A$, a vector in $ \mathbb{R}^{n_A}$.   
\end{Definition}


\begin{Proposition}{\bf (Properties of the Haddamard Product \cite{HornAndJohnson})}
\label{prop:kformulas}
In the following, we implicitly assume that $n_A=n_B$ and $m_A = m_B$ whenever $A \circ B$ is present.
\begin{itemize}
\item[(a)]{\sc Commutivity.}  $A_1 \circ A_2 = A_2\circ A_1$.
\item[(b)]{\sc Scalar Multiplication.}
For any $a_1, a_2 \in \mathbb{R}$,   
$$
(a_1 a_2) (A_1 \circ A_2) = (a_1 A_1) \circ (a_2 A_2). 
$$
\item[(c)] {\sc Distributivity.}
\begin{eqnarray*}
(A_1 + A_2) \circ A_3 &=&  (A_1 \circ A_3)+ (A_2 \circ A_3)
\qquad \mbox{and} \\
A_1 \circ (A_2 + A_3) &=& (A_1 \circ A_2)+ (A_1 \circ A_3).
\end{eqnarray*}
\item[(d)] {\sc Tranposition.}
$
(A_1 \circ A_2)^t = (A_1^t \circ A_2^t).
$
\item[(e)] {\sc Haddamard-Kronecker Distributivity. }
$$
(A_1 \otimes A_2) \circ (A_3 \otimes A_4) = (A_1 \circ A_3) \otimes (A_2 \circ A_4).
$$
\item[(f)] {\sc Diagonal-Kronecker Distributivity.}  When $m_{A_1} = n_{A_1}$ and $m_{A_2} = n_{A_2}$,
$$
\mbox{diag}(A_1 \otimes A_2) = \mbox{diag}(A_1) \otimes \mbox{diag}(A_2).
$$
\end{itemize}

\end{Proposition}

\begin{proof}
(a)-(e) are standard properties. 
For (f), \gedit{ $\mbox{diag}(A_1 \otimes A_2) =$}
\vspace{-0.25cm}
\gedit{
\begin{eqnarray*}
& = &  [ (I_{A_1} \otimes I_{A_2}) \circ (A_1 \otimes A_2) ] ({\bf 1}_{A_1} \otimes {\bf 1}_{A_2}) \\
& = & [ (I_{A_1} \circ A_1) \otimes (I_{A_2} \circ A_2) ] ({\bf 1}_{A_1} \otimes {\bf 1}_{A_2}) \\
& = & [ (I_{A_1} \circ A_1) {\bf 1}_{A_1} ] \otimes [ (I_{A_2} \circ A_2) {\bf 1}_{A_2} ] \\
& = & \mbox{diag}(A_1) \otimes \mbox{diag}(A_2). \qedhere
\end{eqnarray*}
}
\end{proof}


\subsection{Graph Notation}

Let ${\cal G}({\cal V}, {\cal E})$ be a set of $n:=|{\cal V}|$ vertices and $|{\cal E}|$ edges, pair-wise relationships between members of ${\cal V}$ of the form $(i,j) \in {\cal E}$, where $i,j \in {\cal V}$.  We say ${\cal G}$ is {\em undirected} if $(i,j) \in {\cal E}$ implies $(j,i) \in {\cal E}$ for every $(i,j)$ (and ${\cal G}$ is {\em directed} if this doesn't hold for a single edge).   An edge of the form $(i,i) \in {\cal E}$ is a {\em self loop}.

Let $\mathbb{B} = \{0,1\}$.  The matrix $A \in \mathbb{B}^{n \times n}$ is an {\em adjacency matrix} representing ${\cal G}$ if $A_{ij} = 1$ for each $(i,j) \in {\cal E}$ and $A_{ij} = 0$ for each $(i,j) \not\in {\cal E}$.   
Given an adjacency matrix $A$, we use ${\cal G}_A$, ${\cal V}_A$, and ${\cal E}_A$, to represent the associated graph, vertices, and edges, respectively. 
We use a subscript $A$ for many other symbols referring to properties of ${\cal G}_A$ (e.g. $n_A = |{\cal V}_A|$). 

\begin{Remark}
Edge incidence matrices and matrix reorderings are important constructs for the most efficient linear algebraic formulas for performing some actual graph computations \cite{gilbert2015, wolf2017fast, hutchison2017distributed, low2017first}.   However, edge incidence matrices would greatly complicate the Kronecker formula derivations we present, due to their row ordering being arbitrary.   Therefore, we avoid using them in the derivations presented in this work.  
\end{Remark}

\section{Undirected Graphs}
\label{sxn:undir}

Let $A \in \mathbb{B}^{n_A \times n_A}$, $B \in \mathbb{B}^{n_B \times n_B}$, be two {\em adjacency matrices} (possibly with self loops) on $n_A$ and $n_B$ vertices, respectively.
The matrix 
$
C = A \otimes B \in \mathbb{B}^{n_C \times n_C}
$
is an adjacency matrix (possibly with self loops) on $n_C=n_A n_B$ vertices. 
We define index maps 
\begin{eqnarray*}
p(i,k) =  \gamma_{n_B}(k,i), & \quad&
q(j,l) = \gamma_{n_B}(l,j), \\
i(p) = \alpha_{n_B} (p), &\quad&
j(q) = \alpha_{n_B}(q), \\ 
k(p) = \beta_{n_B}(p), &\quad&
l(q) = \beta_{n_B}(q),
\end{eqnarray*}
so 
$C_{p(i,k), q(j,l)} = A_{ij}B_{kl}$
or
$C_{pq} = A_{i(p),j(q)} B_{k(p),l(q)}$.   For diagonal elements, $C_{pp} = A_{i(p),i(p)} B_{k(p),k(p)}$.

\begin{Remark} {\bf (Self-Loops)}
As observed in \cite{Leskovec2010KronFit},\cite{Kepner2018GABB}, putting self loops into the factors of $A$ and $B$ boosts the number of triangles in $C$ significantly.   Therefore we will analyze the case when factors have no self-loops (for simplicity) and cases when one or more of the factors have self loops.   Also, removing all self loops from an adjacency matrix can be written in terms of the Haddamard product, $(C - I_C \circ C)$, making Kronecker product formulas still fairly simple for many types of graph statistics.  The diagonal operator containing only the self-loops, $D_C = I_C \circ C$, is used in many of the following derivations.
\end{Remark}

Throughout this section we provide several formulas for computing exact graph statistics for $C$ that involve Kronecker products of associated graph statistics of $A$ and $B$. The following example provides the reader with sanity checks of the formulas throughout the section.

\begin{Example}
{\bf (Cliques With and Without Self-Loops)} 
\label{ex:cliques}
Let $J_n = {\bf 1}_n{\bf 1}_n^t$ and define the adjacency matrix of a clique of size $n$ as $K_n := J_n - I_n$.   Within the graph associated with $K_n$, the degree of each vertex is $(n-1)$, the 
number of triangles involving each vertex is ${n-1} \choose {2}$, and the number of triangles involving each edge is $(n-2)$.   

Note that $J_n$ is the adjacency matrix of clique of size $n$ where every vertex has a self loop.   We form three simple examples of Kronecker products involving $K_n$ and $J_n$.

\begin{itemize}
\item Ex.~\ref{ex:cliques}(a), no self loops.   Let $C = K_{n_A} \otimes K_{n_B}$.  Then the degree of each vertex is $(n_A n_B +1 -n_A -n_B)$.
The number of triangles involving each vertex is 
$$
\frac{1}{2}(n_A n_B +1 - n_A - n_B)(n_A n_B +4 - 2n_A - 2n_B)
$$
The number of triangles involving each edge is 
$$
(n_A n_B +4 - 2n_A - 2n_B).
$$

\item Ex.~\ref{ex:cliques}(b), self loops in second factor.   Let $C = K_{n_A} \otimes J_{n_B}$.   Then the degree of each vertex is $(n_A n_B  - n_A)$.
The number of triangles involving each vertex is 
$$
\frac{1}{2}(n_A n_B - n_B)(n_A n_B - 2n_B)
$$
The number of triangles involving each edge is 
$$
(n_A n_B - 2 n_B).
$$

\item Ex.~\ref{ex:cliques}(c), self loops in both factors.  Let $ C =  (J_{n_A} \otimes J_{n_B}) - I_{n_C}$, which is equal to $K_{n_C}$.   Then the degree of each vertex is $(n_A n_B-1)$.
The number of triangles involving each vertex is ${n_An_B-1} \choose {2}$ and the 
number of triangles involving each edge is $(n_An_B-2)$.

\end{itemize}

\gedit{Note that Ex.~\ref{ex:cliques}(c) demonstrates that $C$ can have as many triangles as possible, as it is a clique on $n_C$ vertices.}

\end{Example}

\subsection{Degree-Distribution}
\label{sxn:ddist}
As shown in \cite{Leskovec2010KronFit},\cite{Kepner2018GABB}, it is simple to see the degree distribution vector of $C$, $
{\bf d}_C :=  (C - I_C \circ C)  {\bf 1}_C$, in terms of the degree distribution vectors of $A$ and $B$.   Without self loops in $A$ and $B$, $(I_C \circ C) = O_C$, and
$$
{\bf d}_C =  
(A \otimes B) ({\bf 1}_A \otimes {\bf 1}_B) = (  A {\bf 1}_A) \otimes ( B {\bf 1}_B) = {\bf d }_A \otimes {\bf d}_B.
$$
Note that ${\bf d}_C$ is definitely not a perfect power law distribution (for one, no prime greater than max$(n_A, n_B)$ is possible).   However, if $A$ and $B$ have power-law degree distributions (such as a Pareto distribution)
we can estimate the tail behavior of $C$'s degree distribution to be heavy-tailed, as it is a multinomial of heavy-tailed distributions \cite{Leskovec2010KronFit}. However, it is important to note that the ratio of maximum degree to number of nodes is essentially squared,
$$
\frac{\|{\bf d}_C\|_\infty}{n_C} = \left( \frac{\|{\bf d}_A\|_\infty}{n_A}\right)\left( \frac{\|{\bf d}_B\|_\infty}{n_B}\right).
$$
With self-loops in $B$ only, $({\bf d}_C)_p = ({\bf d}_A)_{i(p)}  \left[({\bf d}_B)_{k(p)} +1 \right] $, and with self loops in both factors, 
$$
({\bf d}_C)_p = \left[({\bf d}_A)_{i(p)} + 1 \right]  \left[({\bf d}_B)_{k(p)} +1 \right] - 1.
$$
The squaring of the ratio of maximum degree to the number of nodes is qualitatively different (unless $\|{\bf d}_A\|_\infty \approx n_A-1$ or $\|{\bf d}_B\|_\infty \approx n_B-1$).

\subsection{Triangle Participation of Vertices}

\begin{figure}[t]
   \centering
   \includegraphics[width=1.0in]{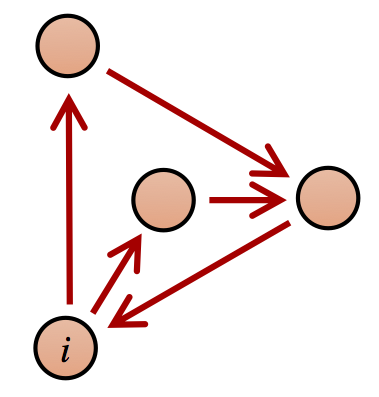} \qquad
   \includegraphics[width=1.0in]{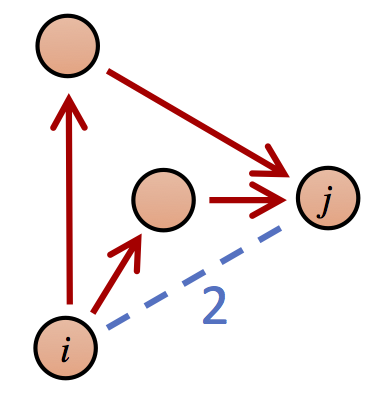} 
   \caption{Triangle participation at vertices and edges in a graph with no self-loops.  Left: $(A^3)_{ii}$ counts the number ways $i$ can take 3 hops to get back to itself, which double counts each triangle, clock-wise (as pictured) and counterclock-wise, so $\frac{1}{2}\mbox{diag}(A^3)$ counts triangles at every vertex.  Right: $(A^2)_{ij}$ counts the number of 2-paths between vertices $i$ and $j$, so $A \circ A^2$ counts triangles at every edge.}
   \label{fig:hadamard}
\end{figure}

\begin{Definition}{\bf (Triangle Participation at Vertices)}
For adjacency matrix $A$, {\em triangle participation at vertices}
is represented by ${\bf t}_{A} \in \mathbb{R}^{n_A}$, a vector that counts the number of undirected triangles at each vertex.   For $A$ undirected with self loops, we have 
$$
{\bf t}_A := \frac{1}{2} \mbox{diag} \Big( (A-A\circ I_A)^3 \Big).
$$ 
Note that when $A$ has no self loops, then ${\bf t}_A = \frac{1}{2} \mbox{diag} ( A^3 )$.
\end{Definition}

\begin{Theorem}{\bf (Triangle Participation at Vertices)}
\label{thm:nsl1}
Let $C = A \otimes B$.   Assume the factors of $C$ are undirected, $A=A^t, B=B^t$, and have no self-loops, 
$\mbox{diag}(A) =  \mbox{diag}(B) = {\bf 0}$.   Then the triangle participation of each vertex is given by
$$
{\bf t}_{C} = 2 {\bf t}_{A} \otimes {\bf t}_{B}. 
$$
\end{Theorem}

\begin{proof}
Using the diag() operator, we have 
\begin{eqnarray*}
{\bf t}_{C} & = & \frac{1}{2} \mbox{diag}(C^3) = \frac{1}{2} \mbox{diag}\left( (A \otimes B) (A \otimes B) (A \otimes B)  \right)
\\&  =&  \frac{1}{2} \mbox{diag}\left( A^3 \otimes B^3 \right)\\
&= & 2 \left( \frac{1}{2} \mbox{diag}(A^3) \right) \otimes \left( \frac{1}{2} \mbox{diag}(B^3) \right) = 2 {\bf t}_{A} \otimes {\bf t}_{B}. \qedhere
\end{eqnarray*}
\end{proof}

One can validate this formula for Ex.~\ref{ex:cliques}(a).  Also, from $\tau(C) = \frac{1}{3} {\bf 1}_C^t {\bf t}_C $ it is easy see that the total number of triangles in $C$ obeys $\tau(C) = 6 \tau(A) \tau(B)$.
Notice that without self-loops in $A$ or $B$ there will always be an even number of triangles for every vertex in ${\cal G}_C$.
More generally, if we allow $B$ to have self loops and $A$ to have none, then $C$ has no self loops.   The formula for triangle participation is still very simple.

\begin{Corollary}{\bf (Self Loops)}
\label{thm:sl1}
Assume the factors of $C = A \otimes B$ are undirected graphs, $A=A^t , B = B^t$, $A$ has no self-loops, 
$\mbox{diag}(A) = {\bf 0}$, but $B$ has self loops, $\mbox{diag}(B) \neq {\bf 0}$.  
Then the triangle participation of each vertex is given by
$$
{\bf t}_{C} = {\bf t}_{A} \otimes \mbox{diag}(B^3). 
\vspace{-.25cm}
$$
\hfill {\footnotesize {Proof in Appendix.}}
\end{Corollary}

Using $\mbox{diag}(J_{n_B}^3) = n_B^2 I_{n_B}$, one can validate this formula for Ex.~\ref{ex:cliques}(b).
Note that in the corollary above $\mbox{diag}(B^3)$ contains in its $k$-th entry double counts of triangles $\mbox{diag}((B- B\circ I)^3)_k$ and the four other three-step sequences from vertex $k$ to itself that involve non-trivial edges and self loops.   For example, if $l$ is connected to $k$, $\mbox{diag}(B^3)_k$ counts the non-triangles
$
\{(k,k), (k,k), (k,k)\},$ $\{(k,k), (k,l), (l,k)\},$ $\{(k,l), (l,k), (k,k)\},$   and  $\{(k,l), (l,l), (l,k)\}.
$

In the fully general case, where $A$ and $B$ both have self loops, we have a more complicated formula.   Let $D_C := I \circ C$. Note that for any $Q \in \mathbb{R}^{n_C \times n_C}$ we have  diag$(D_C Q)  =\mbox{diag}(D_Q D_C) =\mbox{diag}(Q D_C)$,  and $D_C^2 = D_C$ (a diagonal with 0 and 1 entries) to show ${\bf t}_C = \frac{1}{2} \mbox{diag}\left( (C - C\circ I_C)^3 \right) =
$
\begin{eqnarray*}
&& \frac{1}{2} \mbox{diag} \Big(C^3 - C^2 D_C - CD_CC - D_CC^2  \\
&& \quad + \,\, CD_C^2 + D_C C D_C + D_C^2 C - D_C^3 \Big) =\\  
&& \frac{1}{2} \Big( \mbox{diag}(C^3) - 2  \mbox{diag}(C^2 D_C) -\mbox{diag}(C D_C C)  \\
&&  \quad + \,\, 2  \mbox{diag}(D_C^3) \Big) =\\
&& \frac{1}{2} \Big( 
\mbox{diag}(A^3) \otimes \mbox{diag}(B^3) 
- 2 \mbox{diag} (A^2 D_{A}) \otimes \mbox{diag} (B^2 D_{B}) \\
&& \quad  -  \, \,    \mbox{diag} (A D_A A) \otimes \mbox{diag}(B D_B B) \\
&& \quad + \,\, 2 \mbox{diag} (D_A) \otimes \mbox{diag} (D_B)
\Big).
\end{eqnarray*}
One can validate this formula for Ex.~\ref{ex:cliques}(c), using $\mbox{diag}(J_{n}^3) = n^2 I_{n}$, $D_{J_n}=I_n$, ${J_n}^2 = nJ_n$.   

%
%

%
%


\subsection{Triangle Participation of Edges}


\begin{Definition}{\bf (Triangle Participation at Edges)}

{\em Triangle participation at edges}, 
is a $n_A \times n_A$ matrix 
$$ \Delta_A := (A - A \circ I) \circ (A - A \circ I)^2, $$ 
whose $(i,j)$-th entry is the number of triangles in which edge $(i,j)$ participates.   When $A$ has no self-loops, $\Delta_A = (A \circ A^2)$.
\end{Definition}

A useful formula from this definition is ${\bf t}_A = \frac{1}{2} \Delta_A {\bf 1}_A$.

\begin{Theorem}{\bf (Triangle Participation at Edges)}
\label{thm:uedge}
Let $C = A \otimes B$.   Assume the factors of $C$ are undirected, $A=A^t, B=B^t$, and have no self-loops, 
$\mbox{diag}(A) =  \mbox{diag}(B) = {\bf 0}$. The number of triangles in any edge within the graph of $C$ is 
$$
\Delta_C = \Delta_A \otimes \Delta_B.
\vspace{-.5cm}
$$
\hfill {\footnotesize {Proof in Appendix.}}
\end{Theorem}
\begin{Corollary}{\bf (Self Loops)}
Assume the factors of $C$ are undirected, $A=A^t, B=B^t$, and $A$ has no self-loops, 
$\mbox{diag}(A) = {\bf 0}$. The number of triangles in any edge within the graph of $C$ is 
$$
\Delta_C = \Delta_A \otimes (B \circ B^2).
$$
\end{Corollary}
The proof is straightforward because $C \circ I_{C} = O_{C}$, due to $A$ having no self loops.   If $A$ and $B$ both have self loops, then employ $C \circ C = C$, $D_C \circ D_C = D_C^2 = D_C$, $C \circ (C D_C) = (C \circ C) D_C$, and $D_C \circ (C D_C) = D_C$ to see that 
$\Delta_C = (C - D_C) \circ (C- D_C)^2 =$
\begin{eqnarray*}
&= & C \circ C^2 - C\circ (D_C C) - C \circ (C D_C)  \\
&& \quad + \,\, C \circ D_C^2 - D_C \circ C^2 + D_C \circ (D_C C )   \\
&& \quad + \,\, D_C \circ (C D_C) - D_C \circ D_C^2 \\
& = & (A \circ A^2) \otimes (B \circ B^2) - (D_A A) \otimes (D_B B) \\  
&& \quad - \,\, (A D_A ) \otimes (B D_B) + 2 D_A \otimes D_B \\
&& \quad - \,\, (D_A \circ A^2) \otimes (D_B \circ B^2).
\end{eqnarray*}

Again, all three results in this section can be validated by applying the formulas to Ex.~\ref{ex:cliques}(a)-(c).

\subsection{Truss Decomposition}

We discuss deriving simple Kronecker formulas for the $\kappa$-truss and the truss decomposition \cite{Cohen2008ktruss, samsi2017static} of $C = A \otimes B$ and give a simple example showing this is difficult in general.

\begin{Definition}{\bf ($\kappa$-Truss and Truss Decomposition \cite{Cohen2008ktruss, samsi2017static})}
A {\em $\kappa$-truss} in $\mathcal{G}_A$ is a non-trivial, one-component subgraph of such that each edge $(i,j)$ in the $k$-truss participates in at least $(\kappa-2)$ triangles whose edges are all in the $\kappa$-truss.

The {\em truss decomposition} of $\mathcal{G}_A$ is the sequence of edge sets 
$$
{\cal T}_A^{(\kappa)} := \left\{ 
(i,j) \, : \, (i,j) \mbox{ is in a $\kappa$-truss.} 
\right\}
$$ 
for $\kappa=3,4,..., n_A$.   
\end{Definition}

A simple (yet inefficient) algorithm gives the truss decomposition of $A$.  
Set $A' \leftarrow A$.   
Repeat the following for $\kappa=3,...,n_A$, or until there are no more edges. 
Compute $\Delta_{A'}$.
Remove any edge that has less that $(\kappa-2)$ triangles and update $A'$.  
Repeat these edge removal phases for fixed $\kappa$, recomputing $\Delta_{A'}$, removing, and updating $A'$ until no edges are removed.   
Then, set ${\cal T}_A^{(\kappa)}$ equal to all remaining edges in $A'$, increment $\kappa$, and repeat edge removal phases until done.


For $C = A \otimes B$, the formula $\Delta_C = \Delta_A \otimes \Delta_B$ (Thm.~\ref{thm:uedge}) seems useful for mapping the truss decompositions of $A$ and $B$ onto the truss decomposition of $C$.   The following example shows that a simple Kronecker formula is insufficient.

\begin{figure}[t]
   \centering
   \includegraphics[width=3.5in]{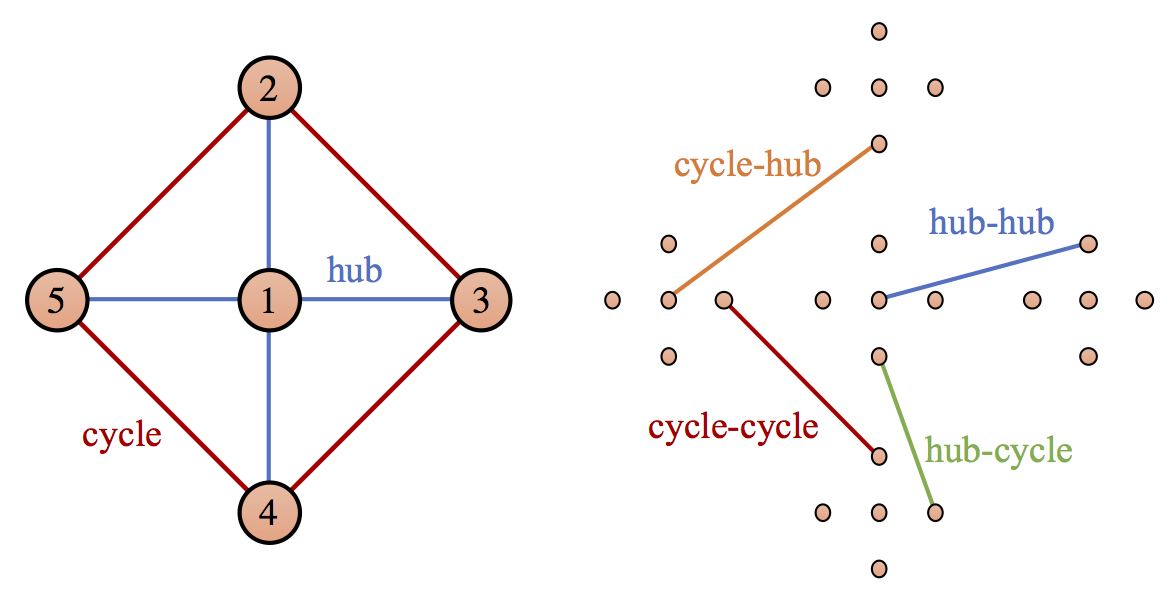} 
   \caption{Graphs from Ex~\ref{ex:truss}. Left: graph associated with $A$ with hub edges colored blue and cycle edges colored red.   Right: vertices from $C = A \otimes A$ with a subset of the edges drawn and labeled as hub-hub (blue), hub-cycle (green), cycle-hub (orange), and cycle-cycle (red).}
   \label{fig:trussex}
\end{figure}

\begin{Example}
\label{ex:truss}
Let $A$ be a 4-cycle with an added hub, $A = K_5 - {\bf e}_2 {\bf e}_4^t - {\bf e}_4 {\bf e}_2^t - {\bf e}_3 {\bf e}_5^t - {\bf e}_5 {\bf e}_3^t$, a graph with 5 vertices, 8 edges, and 4 triangles (see Fig.~\ref{fig:trussex}).   
The {\em cycle} edges of $A$ (those not involving vertex 1) all participate in a single triangle, whereas the {\em hub} edges all participate in 2 triangles.   
All edges from the graph of $A$ are in the 3-truss, yet no edges are in the 4-truss.  

Let $C = A \otimes A$, which has an associated graph with 25 vertices, 128 edges, and 96 triangles.    
Due to the Kronecker decomposition of $C$, edges can be described as combinations of hub and cycle edges (see the right side of Fig.~\ref{fig:trussex}).   
Using the Kronecker formula (Thm.~\ref{thm:uedge}) for triangle participation of edges, we see there are 32 edges that participate in 1 triangle (cycle-cycle edges), 64 that participate in 2 triangles (hub-cycle and cycle-hub edges), and 32 that participate in 4 triangles (hub-hub edges).   
The graph of $C$ has 128 edges in the 3-truss, 80 edges in the 4-truss, and zero in the 5-truss, more complicated structure than that of a simple Kronecker product.      
\end{Example}

For a Kronecker formula for the truss decomposition of $C = A \otimes B$, either sophisticated assumptions on $A$ and $B$ need to be made or diagnostics of the intermediate phases of the truss decomposition algorithm need to be involved in the formula.   In this work, we make fairly strong assumptions on one factor (edges of $B$ participate in at most one triangle) that imply a simple formula for the truss decomposition of $C$.    Note that edges of $B$ participate in at most one triangle is a stronger assumption than  ${\cal T}_B^{(3)}$ being the only nontrivial set of the truss decomposition of $B$.

\begin{Theorem}{\bf (Truss Decomposition)}
Let $C = A \otimes B$.   Assume the factors of $C$ are undirected, $A=A^t, B=B^t$, have no self-loops, 
$\mbox{diag}(A) =  \mbox{diag}(B) = {\bf 0}$, and edges of $B$ participate in no more than one triangle, or $\Delta_B \leq 1$.   We have,
$(p,q) \in {\cal T}^{(\kappa)}_C$ if and only if 
$$
(i, j) \in  {\cal T}_A^{(\kappa)}
\quad \mbox{and} \quad
(k, l) \in {\cal T}_B^{(3)},
$$
for $i = \alpha_{n_B}(p), j = \alpha_{n_B}(q),  k = \beta_{n_B}(p), l  = \beta_{n_B}(q).$
\end{Theorem}
\begin{proof}
If $(\Delta_B)_{k,l}$ is zero, then $(p,q)$ is in no triangle and is in no $\kappa$-truss. Let ${\cal E}'_C$ be the set of edges $(p,q)$ for which $(\Delta_B)_{k,l}=1$.
Due to the strong assumptions on $B$, the simple truss decomposition algorithm applied to the remaining edges in ${\cal E}'_C$ proceeds in lock step with that of ${\cal G}_A$ in the sense that any edge $(p,q)$ is removed at phase $\phi$ if and only if $(i,j)$ is also removed at phase $\phi$, as $(\Delta_C)_{p,q} = (\Delta_A)_{i,j}$.
\end{proof}
Note that most real-world graphs do not satisfy the assumptions on $B$ made in the previous result.   Until more sophisticated theory weakens these assumptions, we have two possible strategies for generating such $B$ that are scale-free.
\begin{itemize}
\item[(a)] Delete edges in a real-world graph until all edges participate in at most one triangle, while maintaining connectivity (with any spanning tree).
\item[(b)] Use a simple graph generator  (based on preferential attachment \cite{Barabasi1999}) to yield power-law graphs where each edge participates in no more than one triangle.    The generator starts with a single edge and  proceeds as follows.   For each new node $u$, pick edge $(i,j)$ uniformly at random from the previously existing edges.  Pick vertex $v$ from $\{i,j\}$ uniformly at random and add $(u,v)$ to the list of edges.  If the number of triangles that $(i,j)$ participates in is zero, then let $w$ be vertex in $\{i,j\}$ that wasn't already attached, add $(u,w)$ to the list of edges, and increment the triangle count for $(i,j), (u,v),$ and $(u,w)$.  Repeat for a new $u$ until the desired number of vertices is met.
\end{itemize}

\section{Directed Graphs}
\label{sxn:dir}

\subsection{Directed and Reciprocal Parts}

We use the most general directed graph model, where directed edges and reciprocal edges are treated differently \cite{SePiDuKo16}.    Under this model topological diversity is more locally detectable than the standard model.   For example, a vertex that has 3 reciporical edges is very different than a vertex that has 6 disjoint edges (3 incoming and 3 outgoing). 

\begin{Definition}{\bf (Reciprocal and Directed Edges)}
Given a directed graph ${\cal G}({\cal V}, {\cal E})$, we divide the edges in two disjoint sets, ${\cal E} = {\cal E}_d \cup {\cal E}_r$, where ${\cal E}_d$ contains all {\em directed edges} and ${\cal E}_r$ contains all {\em reciprocal edges}.
If $(i,j) \in {\cal E}$ and $(j,i) \not \in {\cal E}$ then $(i,j) \in {\cal E}_d$.
If $(i,j) \in {\cal E}$ and $(j,i) \in {\cal E}$ then $(i,j) \in {\cal E}_r$.
\end{Definition}

\begin{Definition}{ \bf (Reciprocal and Directed Parts of $A$)}
We linearly decompose a non-symmetric adjacency matrix $A$ into 
$$
A = A_r + A_d, 
\qquad
A_r := A^t  \circ A, 
\quad
A_d := A - A_r,
$$
where $A_r$ is the {\em reciprocal part} of $A$ and $A_d$ is the {\em directed part} of $A$.   Also, the {\em undirected version} of  $A$ is  $A_u := A + A_d^t$. 
\end{Definition}

For $B$ and $C = A \otimes B$, we have the respective decompositions,  $B= B_r + B_d$, and $C = C_r + C_b$.   Moreover, $C_r$ and $C_d$ have simple formulas in terms the respective parts of $A$ and $B$,
\begin{eqnarray*}
C_r & = & (A \otimes B)^t \circ (A \otimes B)  = (A^t \otimes B^t) \circ (A \otimes B) \\& = &(A^t \circ A) \otimes (B^t \circ B)
 =  A_r \otimes B_r\\
\\
C_d & = & C - C_r = (A \otimes B)  -  (A_r \otimes B_r) \\ & = & (A_r + A_d) \otimes (B_r + B_d)  -  (A_r \otimes B_r)\\
& = & A_d \otimes B_r + A_r \otimes B_d + A_d \otimes B_d. 
\end{eqnarray*}

These equations could be used for general $A$ and $B$.   We take the approach of restricting $B = B^t$ (associated with an undirected graph) so every edge in $B$ is reciprocal to greatly simplify the Kronecker formulas we derive, as  
$$
B_d = O,
\quad
\Longrightarrow
\quad
C_r = A_r \otimes B, \qquad C_d = A_d \otimes B.
$$
In cases where the input data for the second factor is directed, we advocate using the undirected version for $B$ so the formulas remain simple.

\subsection{Degree-Distribution}
We have simple formulas for the standard in/out degree vectors, ${\bf d}_C^{out} :=  C  {\bf 1}_C$, ${\bf d}_C^{in} :=  C^t  {\bf 1}_C$,
$$
{\bf d}_C^{out} =  (A \otimes B) ({\bf 1}_A \otimes {\bf 1}_B) = (  A {\bf 1}_A) \otimes ( B {\bf 1}_B) = {\bf d}^{out}_A \otimes {\bf d}^{out}_B.
$$
$$
{\bf d}_C^{in} =  (A \otimes B)^t ({\bf 1}_A \otimes {\bf 1}_B) = (  A^t {\bf 1}_A) \otimes ( B^t {\bf 1}_B) = {\bf d}^{in}_A \otimes {\bf d}^{in}_B.
$$
\gedit{Under the directed/reciprocal edge model and assuming $B= B^t$, we also have similar formula for reciprocal, directed-out, and directed-in degree vectors, 
${\bf d}_{C_r} =  {\bf d}_{A_r} \otimes {\bf d}_B$,
${\bf d}_{C_d}^{out} = {\bf d}^{out}_{A_d} \otimes {\bf d}_B$,
and ${\bf d}_{C_d}^{in} = {\bf d}^{in}_{A_d} \otimes {\bf d}_B$.}


\subsection{Directed Triangle Participation of Vertices}


\begin{figure}[t]
   \centering
   \includegraphics[width=3.5in]{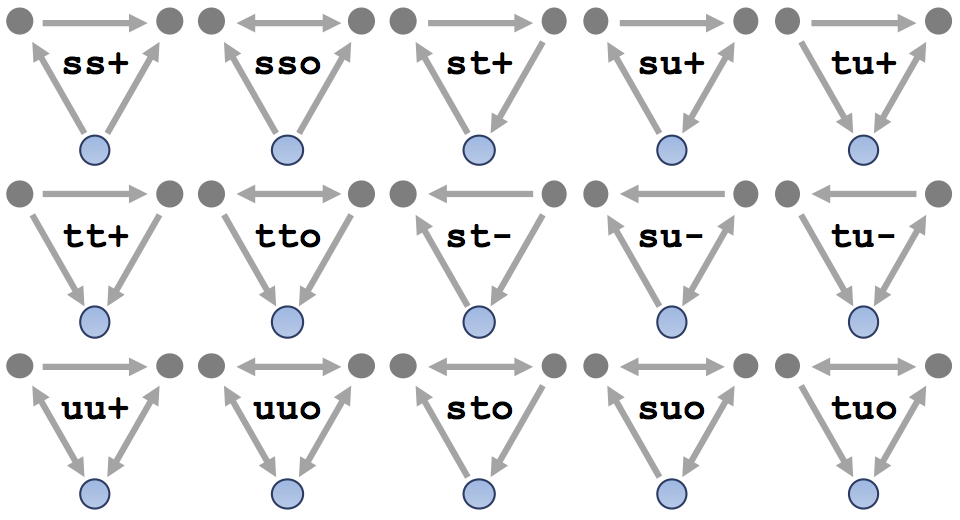} 
   \caption{Fifteen possible different combinations (removing symmetries) of directed triangles with reciprocal and directed edges from a vertex's perspective. \gedit{The triangle types are depicted by listing first the central vertex's role in the triangle (is it a source 's', target 't', or undirected 'u' with respect to the two incident edges).   The last character represents the direction of the remaining edge, oriented ('+' forward, '-' backward, or 'o' undirected) in listed order.} }
   \label{fig:vtris}
\end{figure}

\begin{Definition}{\bf (Directed Triangle Participation at Vertices)}
Directed triangle participation of type $\uptau$ at vertices, ${\bf t}_{A}^{(\uptau)} \in \mathbb{R}^{n_A}$, is a vector that counts the number of directed triangles of type $\uptau$ at each vertex (see Fig.~\ref{fig:vtris}).
For $A$ directed with no self loops, we have 
$$
\begin{array}{|c|c|c|}
\hline
{\bf t}^{(ss+)}_A = &
{\bf t}^{(sso)}_A =  &
{\bf t}^{(ss-)}_A = \\
\mbox{diag}(A_d^t  A_d^2), &
\frac{1}{2}\mbox{diag}(A_d^t A_r A_d), &
{\bf t}^{(ss+)}_A, \\
\hline 
{\bf t}^{(su+)}_A =&
{\bf t}^{(suo)}_A =&
{\bf t}^{(su-)}_A = \\
\mbox{diag}(A_r A_d^2), &
\mbox{diag}(A_r^2 A_d), &
\mbox{diag}(A_r A_d^t A_d), \\
\hline
{\bf t}^{(st+)}_A =&
{\bf t}^{(sto)}_A =&
{\bf t}^{(st-)}_A =\\
\mbox{diag}(A_d^3), &
\mbox{diag}(A_d A_r A_d), &
\mbox{diag}(A_d A_d^t A_d), \\
\hline
{\bf t}^{(us+)}_A  =  {\bf t}^{(su-)}_A&
{\bf t}^{(uso)}_A  =  {\bf t}^{(suo)}_A &
{\bf t}^{(us-)}_A  =  {\bf t}^{(su+)}_A \\
\hline
{\bf t}^{(uu+)}_A =&
{\bf t}^{(uuo)}_A =&
{\bf t}^{(uu-)}_A = \\
 \mbox{diag}(A_r A_d A_r) &
 \frac{1}{2} \mbox{diag}(A_r^3)  &
 {\bf t}^{(uu+)}_A \\
 \hline
{\bf t}^{(ut+)}_A =&
{\bf t}^{(uto)}_A =&
{\bf t}^{(ut-)}_A = \\
\mbox{diag}(A_d^2 A_r), &
\mbox{diag}(A_d A_r^2), &
\mbox{diag}(A_d A_d^t A_r), \\
\hline
{\bf t}^{(ts+)}_A  =  {\bf t}^{(st-)}_A&
{\bf t}^{(tso)}_A  =  {\bf t}^{(sto)}_A &
{\bf t}^{(ts-)}_A  =  {\bf t}^{(st+)}_A \\
\hline
{\bf t}^{(tu+)}_A = {\bf t}^{(ut-)}_A&
{\bf t}^{(tuo)}_A = {\bf t}^{(uto)}_A &
{\bf t}^{(tu-)}_A = {\bf t}^{(ut+)}_A \\
\hline
{\bf t}^{(tt+)}_A =&
{\bf t}^{(tto)}_A =&
{\bf t}^{(tt-)}_A = \\
\mbox{diag}(A_d (A_d^t)^2), &
 \frac{1}{2}\mbox{diag}(A_d A_r A_d^t ), &
{\bf t}^{(tt+)}_A. \\
\hline
\end{array}
$$
\end{Definition}

When $B$ has no reciprocal edges and $A$ has no self loops, we have a simple formula for each flavor $\uptau$ of directed directed triangle in $C$ at vertex $p$ based on the count of the same flavor of triangle in $A$ and undirected triangle count with $B$, at the associated vertices $i(p)$ in $A$'s graph and $k(p)$ in $B$'s graph.

\begin{Theorem}{\bf (Directed Triangle Participation at Vertices)}
\label{thm:dverts}
Let $C = A \otimes B$.   Assume the right factor of $C$ is undirected, $B_d = O$, and $A$ has no self-loops, 
$\mbox{diag}(A) ={\bf 0}$.   
For every type of directed triangle $\uptau$, we have 
$$
{\bf t}^{(\uptau)}_{C} = {\bf t}^{(\uptau)}_{A} \otimes \mbox{diag} (B^3).
\vspace{-0.5cm}
$$
\hfill {\footnotesize {Proof in Appendix.}}
\end{Theorem}



\subsection{Directed Triangle Participation of Edges}

\begin{figure}[t]
   \centering
   \includegraphics[width=3.5in]{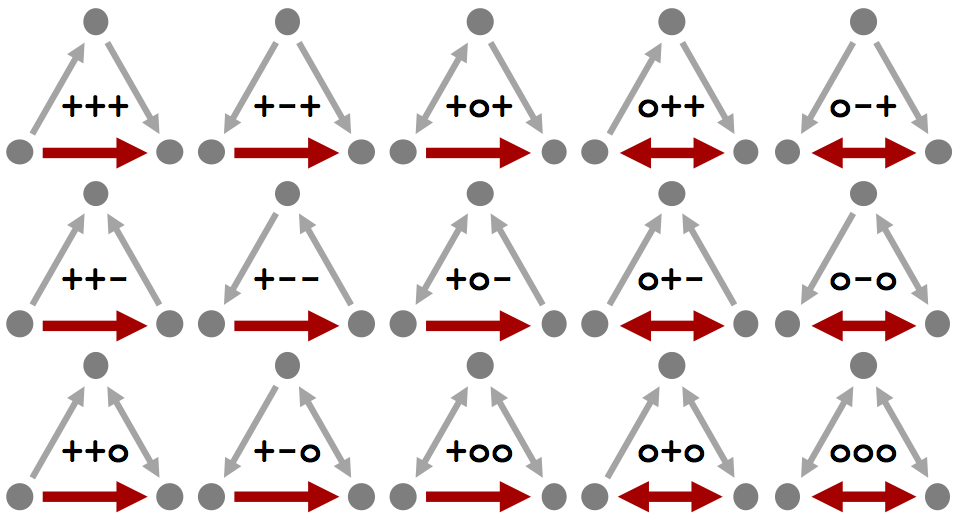} 
   \caption{Fifteen possible different combinations (removing symmetries) of directed triangles with reciprocal and directed edges from an edge's perspective. 
   \gedit{The triangle types are depicted by listing first the central edge's role in the triangle (is it directed '+' or not 'o').   The last two characters represent the directionality of the remaining edges, oriented ('+' forward, '-' backward, or 'o' undirected) in listed order.}
   } 
   \label{fig:etris}
\end{figure}

\begin{Definition}{\bf (Directed Triangle Participation at Edges)}
Directed triangle participation of type $\uptau$ at edges, $\Delta_{A}^{(\uptau)} \in \mathbb{R}^{n_A \times n_A}$, is a matrix that counts the number of directed triangles of type $\uptau$ at each edge (see Fig.~\ref{fig:etris}).
For $A$ directed with no self loops, we have 
$$
\begin{array}{|c|c|c|}
\hline
\Delta^{(+++)}_A = &
\Delta^{(++-)}_A =  &
\Delta^{(++o)}_A = \\
A_d \circ  (A_d^2), &
A_d \circ (A_d^t A_d), &
A_d \circ (A_rA_d), \\
\hline 
\Delta^{(+-+)}_A = &
\Delta^{(+--)}_A =  &
\Delta^{(+-o)}_A = \\
A_d \circ  (A_d A_d^t), &
A_d \circ (A_d^t)^2, &
A_d \circ (A_rA_d^t), \\
\hline 
\Delta^{(+o+)}_A = &
\Delta^{(+o-)}_A =  &
\Delta^{(+oo)}_A = \\
A_d \circ  (A_d A_r), &
A_d \circ (A_d^tA_r), &
A_d \circ (A_r^2), \\
\hline
\Delta^{(o++)}_A = &
\Delta^{(o+-)}_A =  &
\Delta^{(o+o)}_A = \\
A_r \circ  (A_d^2), &
A_r \circ (A_d^tA_d), &
A_r \circ (A_r A_d), \\
\hline
\Delta^{(o-+)}_A = &
\Delta^{(o--)}_A =  &
\Delta^{(o-o)}_A = \\
A_r \circ  (A_d A_d^t), &
\Delta^{(o++)}_A, &
A_r \circ (A_r A_d^t), \\
\hline
\Delta^{(oo+)}_A = &
\Delta^{(oo-)}_A =  &
\Delta^{(ooo)}_A = \\
\Delta^{(o+o)}_A, &
\Delta^{(o-o)}_A, &
A_r \circ (A_r^2), \\
\hline
\end{array}
$$
\end{Definition}


\begin{Theorem}{\bf (Directed Triangle Participation at Edges)}
\label{thm:dedges}
Let $C = A \otimes B$.   Assume the right factor of $C$ is undirected, $B_d = O$, and $A$ has no self-loops, 
$\mbox{diag}(A)= {\bf 0}$.   
For every type of directed triangle $\uptau$, we have 
$$
\Delta^{(\uptau)}_{C} = \Delta^{(\uptau)}_{A} \otimes \left[B \circ(B^2)\right].
\vspace{-.25cm}
$$
\hfill {\footnotesize {Proof in Appendix.}}
\end{Theorem}


One can further generalize the results in this section to the cases $B \neq B^t$ and diag$(A) \neq {\bf 0}$, yet the Kronecker formulas have many terms and are beyond the scope of this paper.

\section{Vertex-Labeled Graphs}
\label{sxn:lab}

\begin{figure}[t]
   \centering
   \includegraphics[width=3.5in]{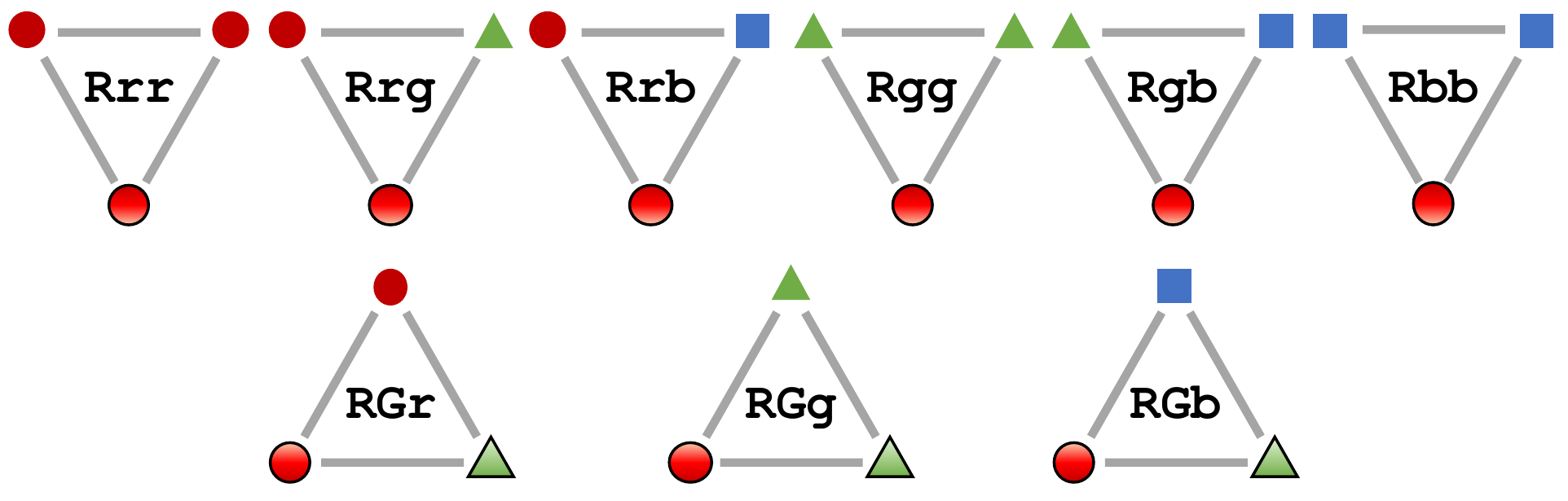} 
   \caption{\gedit{Given the label of a vertex, there are ${|{\cal L}|+1} \choose {2}$ types of triangles the vertex can participate in, removing symmetries.   Given the labels of the two vertices of an edge, there are $|{\cal L}|$ types of triangles an edge can participate in.  The $|{\cal L}|=3$ case is shown.   Triangles types $(q_1,q_2,q_3)$ are depicted with capital letters  representing the label(s) of the central vertex/edge and the lowercase letters representing the labels of the other vertice(s). }
   } 
   \label{fig:ltris}
\end{figure}

Let ${\cal G}({\cal V}, {\cal E}, {\cal L}, f(.))$ be a {\em vertex-labeled} graph, which is a typical graph with a set of labels ${\cal L}:=\{1,2,..., |{\cal L}|\}$ for which the function $f$ maps every vertex into ${\cal L}$.  We often refer to label as the {\em color} of the vertex.   In our description here, we assume $|{\cal L}|=3$ and call color 1 {\em red}, color 2 {\em green}, and color 3 {\em blue} (as in Fig.~\ref{fig:ltris}).   With more labels there are more types of triangles.  To be able to express edges involving only certain labels in linear algebra, we introduce projection operators that filter based on labels.

\begin{Definition}{\bf (Label Filters)}
Let ${\cal G}_A$ be a vertex-labeled graph.   Let ${\bf e}_i$ to be the $i$-th canonical basis vector of dimension $n_A$.
Define the {\em $q$-th label filter}, to be the projection onto all vertices $i$ with label $f_A(i) = q$, or
$$ 
\Pi_{A, q} = \sum_{i \, :\, f_A(i) = q} {\bf e}_i { \bf e}_i^t,
$$
\end{Definition}

The matrix $\Pi_{A, 2} A \Pi_{A, 1}$ is nonzero only for edges from a color 1 vertex to a color 2 vertex.  The number of paths with a certain color sequence is counted easily with filtered matrix multiplication. For example, the $(i,j)$-th entry of matrix $\Pi_{A,3} A \Pi_{A,2} A \Pi_{A,1}$ stores the number of paths of length two starting at a color 1 vertex $j$ going through any color 2 vertex and ending on a color 3 vertex $i$.

\begin{Definition}{\bf (Labeled Triangle Participation at Vertices)}
Labeled triangle participation of type $\uptau = (q_1,q_2,q_3)$ at vertices, ${\bf t}_{A}^{(\uptau)} \in \mathbb{R}^{n_A}$, is a vector that counts the number of labeled triangles of type $\uptau$ at each vertex (see Fig.~\ref{fig:ltris}).
For $A$ with no self loops, if $q_2 = q_3$, we have 
$$
{\bf t}_{A}^{(\uptau)} = 
\frac{1}{2} \mbox{diag} \left( 
\Pi_{A,q_1} A \Pi_{A,q_3} A \Pi_{A,q_2} A \Pi_{A,q_1} 
\right),
$$
whereas,  if $q_2 = q_3$, we have 
$$
{\bf t}_{A}^{(\uptau)} = 
\mbox{diag} \left( 
\Pi_{A,q_1} A \Pi_{A,q_3} A \Pi_{A,q_2} A \Pi_{A,q_1} 
\right),
$$
\end{Definition}

Assume ${\cal G}_A$ is labeled, undirected, with no self loops and ${\cal G}_B$ is unlabeled and undirected. Let $C = A \otimes B$ and let ${\cal G}_C$ be labeled, with the labels inherited from directly from ${\cal G}_A$, $f_C(p) := f_A(\alpha_{n_B}(p))$.   Under this construction, $\Pi_{C,q} = \Pi_{A,q} \otimes I_B$, making it fairly straightforward to derive Kronecker formulas for labeled triangle participation. 

\begin{Theorem}{\bf (Labeled Triangle Participation at Vertices)}
\label{thm:lverts}
Let $C = A \otimes B$.   Assume ${\cal G}_B$ is unlabeled and $A$ has no self-loops, 
$\mbox{diag}(A) ={\bf 0}$.   
For every type of labeled triangle $\uptau = (q_1,q_2,q_3)$, we have 
$$
{\bf t}^{(\uptau)}_{C} = {\bf t}^{(\uptau)}_{A} \otimes \mbox{diag} (B^3).
\vspace{-.25cm}
$$
\hfill {\footnotesize {Proof in Appendix.}}
\end{Theorem}

\begin{Definition}{\bf (Labeled Triangle Participation at Edges)}
Labeled triangle participation of type $\uptau = (q_1,q_2,q_3)$ at edges, $\Delta_{A}^{(\uptau)} \in \mathbb{R}^{n_A \times n_A}$, is a matrix that counts the number of labeled triangles of type $\uptau$ at each edge (see Fig.~\ref{fig:ltris}).
For $A$ with no self loops, we have 
$$
\Delta_{A}^{(\uptau)} = (\Pi_{A,q_2} A \Pi_{A,q_1}) \circ (A \Pi_{A,q_3} A).
$$
\end{Definition}

\begin{Theorem}{\bf (Labeled Triangle Participation at Edges)}
\label{thm:ledges}
Let $C = A \otimes B$.   Assume ${\cal G}_B$ is unlabeled and $A$ has no self-loops, 
$\mbox{diag}(A) ={\bf 0}$.   
For every type of labeled triangle $\uptau = (q_1,q_2,q_3)$, we have 
$$
\Delta^{(\uptau)}_{C} = \Delta^{(\uptau)}_{A} \otimes [B \circ (B^2)].
$$
\hfill {\footnotesize {Proof in Appendix.}}
\end{Theorem}


\begin{figure}[t]
   \centering
   \includegraphics[width=2.75in]{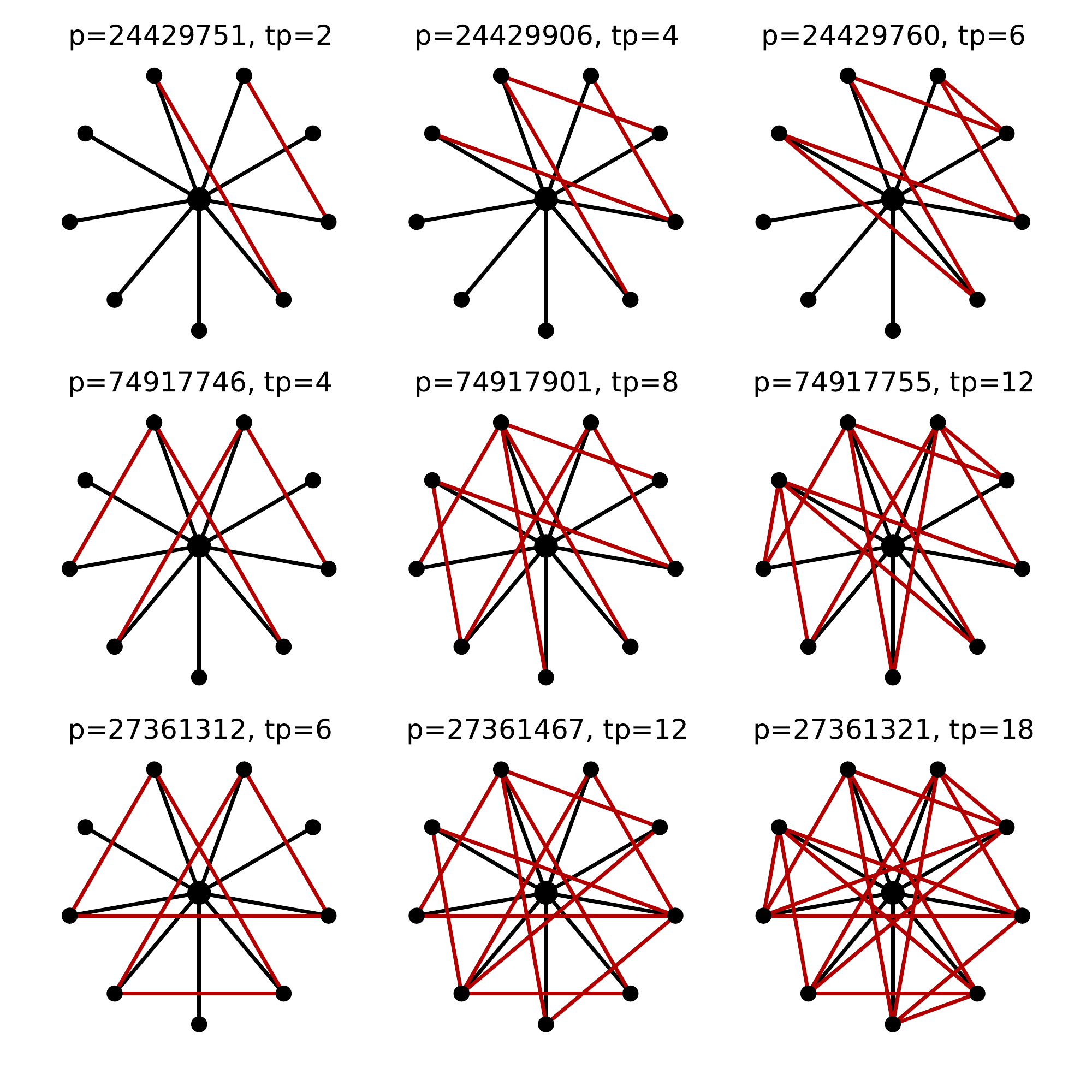} \\
   \hrule

   \includegraphics[width=2.75in]{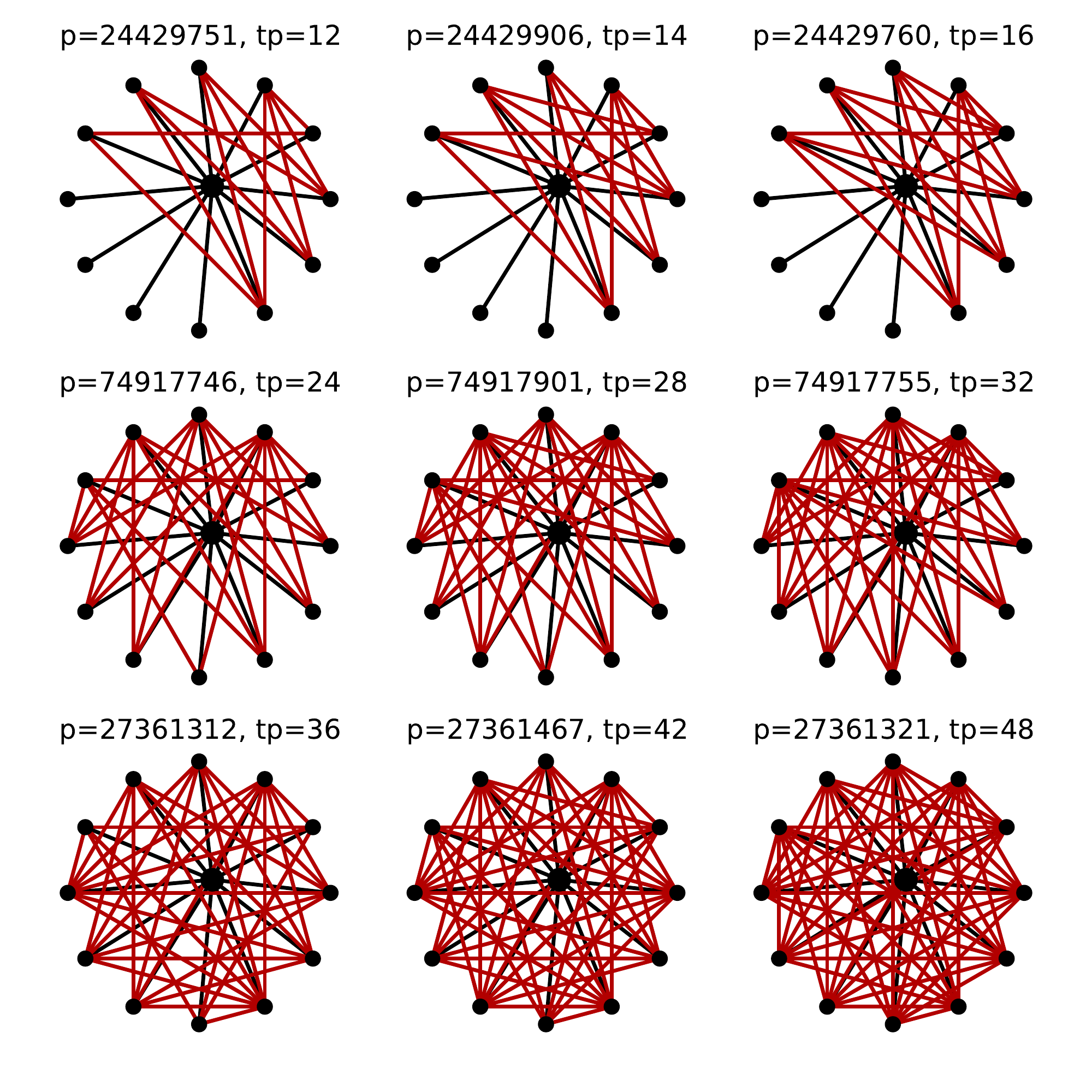}
   \caption{\gedit{Egonets of 9 different vertices for two Kronecker product graphs $A \otimes A$ and $A \otimes B$ built from $A$ and $B = A+I$ derived from the undirected self-loop-free version of the {\tt web-NotreDame} graph.  Three degree-3 vertices in ${\cal G}_A$ with 1, 2, and 3 triangles were selected, in turn yielding 9 vertices associated vertices in ${\cal G}_C$. Top: $A \otimes A$, where the vertices are all degree-9.   Bottom: $A \otimes B$ having no self loops, where the vertices are all degree-12 (both agree with the degree distribution formulas).   The number of triangles present at each vertex agrees with the results in Thm.~\ref{thm:nsl1} and Cor.~\ref{thm:sl1}.}}
   \label{fig:enets}
\end{figure}

\section{Experiments}

We verified the results in this paper on several small examples where $A$ and $B$ had hundreds or thousands of vertices and edges, both by building $C$ entirely and explicitly checking the triangle statistics at each vertex and by constructing individual egonets \gedit{(induced subgraphs of vertex neighborhoods)} of vertices in $C$ and comparing the local triangle statistics to those prescribed by the Kronecker formulas.


\gedit{
We use the undirected version of the {\tt web-NotreDame} graph to verify formulas from Thm.~\ref{thm:nsl1} and Cor.~\ref{thm:sl1}.   
First we removed all self loops in the original data.
The matrices $A$ and $B$ were both derived from the same 325,729 vertex, 1,090,108 edge graph with 4,308,495 triangles: $A$ has no self-loops and $B = A+I$ has a self loops at every vertex.   
We consider $A \otimes A$, which has 106,099,381,441 vertices, 2,376,670,903,328 edges, and 111,378,774,990,150 triangles, and
$A \otimes B$, which has 106,099,381,441 vertices, 2,731,750,692,060 edges, and 111,378,774,990,150 triangles.   The hundred-trillion triangle counts were computed via Kronecker formulas in about 10.5 seconds on a commodity laptop by applying the algorithm from \cite{Chiba1985} to $A$, utilizing 7,734,429 wedge checks (nearly square root complexity of the counts themselves). 
The various graph statistics are summarized in the following table.
\begin{center}
\begin{tabular}{|c||c|c|c|} 
\hline
Matrix & Vertices & Edges & Triangles \\
\hline
$A$ & 325.7K & 1.1M & 4.3M \\
$B = A + I$ & 325.7K & 1.4M & 4.3M \\
$A\otimes A$ & 106.1B & 2.38T & 111.4T \\
$A\otimes B$ & 106.1B & 2.73T & 141.0T \\ 
\hline
\end{tabular}
\end{center}
The triangle participation agreed with Thm.~\ref{thm:nsl1}, Cor.~\ref{thm:sl1}, and Thm.~\ref{thm:uedge} at every vertex and edge we sampled and built an egonet (on a commodity laptop).
From the original graph (with 1 added to every vertex id), we picked three vertices $\{76,231,85\}$  that each had degree 3 and respectively 1, 2, and 3 triangles and then plotted the egonets of the nine vertices that they correspond to in the Kronecker product graph (See Fig.~\ref{fig:enets}).
This visually confirms that the vertex triangle triangle participation for $C$ matches the theory exactly.
}





 \section{Conclusion}
 This work has extended nonstochastic Kronecker product graph generator ability to cheaply produce exact results of triangle statistics during the generation phase by deriving several Kronecker-based formulas that relate triangle statistics of a large Kronecker graph to graph statistics of the smaller factors.
 This suggests nonstochastic Kronecker product graph generators are an extremely attractive benchmark for many combinatorial graph analytics, allowing researchers to test their implementations on stable and reproducible problems with ground truth solution. 
 Further theoretical extensions of this type are likely to make non-stochastic Kronecker product graph generators useful in the same regard for exotic pattern matching, shortest path algorithms, and many other  challenging large-scale distributed graph computations.

 \appendix

\subsection{Proofs of Selected Results}
\begin{proof}[Proof of Cor.~\ref{thm:sl1}]
Because $A\circ I_A = O_A$, we have $C \circ I = (A \circ I_A) \otimes (B \circ I_B) = O_C$.   Then
\begin{eqnarray*}
{\bf t}_C & = & \frac{1}{2} \mbox{diag}\left( (C - C\circ I_C)^3 \right)  = \frac{1}{2} \mbox{diag}(C^3) \\
& = & \left(\frac{1}{2}  \mbox{diag}(A^3) \right) \otimes \mbox{diag}(B^3) = {\bf t}_{A} \otimes \mbox{diag}(B^3). \qedhere
\end{eqnarray*}
\end{proof}
\vspace{-.25cm}
\begin{proof}[Proof of Thm.~\ref{thm:uedge}]
\begin{eqnarray*}
\Delta_C & = & (C \circ C^2) = \Big([A \otimes B] \circ [(A \otimes B )^2]\Big) \\ & =&  \Big([A \otimes B] \circ [(A^2 \otimes B^2 )]\Big) \\
& = & \Big([A \circ A^2] \otimes [B \circ B^2]\Big) =  \Delta_A \otimes \Delta_B. \qedhere
\end{eqnarray*}
\end{proof}
\vspace{-.25cm}
\begin{proof}[Proof of Thm~\ref{thm:dverts}]
$A \circ I_A = O_A$ implies $C \circ I_c = O_C$.
A prototypical case that is double counted due to symmetry, $\uptau = (sso)$,  
$\frac{1}{2} \mbox{diag} (C_d C_r C_d^t) = $
\begin{eqnarray*}
& = & \frac{1}{2} \mbox{diag} \left( 
(A_d \otimes B)
(A_r \otimes B)
(A_d^t \otimes B) 
\right) \\
& = & \frac{1}{2} \mbox{diag} \left( 
(A_d A_r A_d^t \otimes B^3)
\right) \\
& = & 
\frac{1}{2} \left[
(A_d A_r A_d^t \otimes B^3) \circ (I_A \otimes I_B) 
\right] ({\bf 1}_A \otimes {\bf 1}_B)  \\
& = & 
\left[
\frac{1}{2} [ ((A_d A_r A_d^t) \circ I_A )  {\bf 1}_A] \otimes 
[ ((B^3) \circ I_B )  {\bf 1}_B] 
\right] \\
& = & 
 \left[
\frac{1}{2} \mbox{diag} (A_d A_r A_d^t) \otimes 
 \mbox{diag} (B^3)
\right],
\end{eqnarray*}
and a prototypical case that is not double counted, $\uptau' = (st+)$,
\begin{eqnarray*}
\mbox{diag} (C_d^3) 
& = &  \mbox{diag} \left( 
(A_d \otimes B)
(A_d \otimes B)
(A_d \otimes B) 
\right) \\
& = & \mbox{diag} \left( 
(A_d^3 \otimes B^3)
\right) \\
& = & 
\left[
(A_d^3 \otimes B^3) \circ (I_A \otimes I_B) 
\right] ({\bf 1}_A \otimes {\bf 1}_B)  \\
& = & 
[ ((A_d^3) \circ I_A )  {\bf 1}_A] \otimes 
 [ ((B^3) \circ I_B )  {\bf 1}_B] 
\\
& = & 
\mbox{diag} (A_d^3) \otimes 
 \mbox{diag} (B^3)
. \qedhere
\end{eqnarray*}
\end{proof}
\vspace{-.25cm}
\begin{proof}[Proof of Thm~\ref{thm:dedges}]
We show this for a prototypical case: 
\begin{eqnarray*}
\Delta_C^{(+-o)} & = & C_d \circ (C_r C_d^t) \\ 
&=& (A_d \otimes B) \circ \left[ (A_r \otimes B)(A_d^t \otimes B) \right] \\
&=& (A_d \otimes B) \circ \left[ A_r A_d^t \otimes B^2\right] \\
& = & \left[ A_d \circ (A_rA_d) \right] \otimes \left[ B \circ (B^2)\right]\\
& = &  \Delta_A^{(+-o)} \otimes \left[ B \circ (B^2)\right]. \qedhere
\end{eqnarray*}
\end{proof}
 \vspace{-.25cm}
\begin{proof}[Proof of \ref{thm:lverts}]
We only show the $q_2 \neq q_3$ case: ${\bf t}^{(q_1,q_2,q_3)}_{C} = $
\begin{eqnarray*}
 & = &  \mbox{diag}\left( 
\Pi_{C,q_1} C \Pi_{C,q_3} C \Pi_{C,q_2} C \Pi_{C,q_1} 
\right)\\
& = & \mbox{diag}\Big( ( \Pi_{A,q_1} \otimes I_B) (A \otimes B) \\
&& \qquad * \,\, (( \Pi_{A,q_3} \otimes I_B) (A \otimes B) \\
&& \qquad * \,\, ( \Pi_{A,q_2} \otimes I_B) (A \otimes B)( \Pi_{A,q_1} \otimes I_B) \Big) \\
& = & \mbox{diag}(\Pi_{A,q_1} A \Pi_{A,q_3} A \Pi_{A,q_2} A \Pi_{A,q_1} ) \otimes  \mbox{diag}(B^3) \\
& = & {\bf t}^{(q_1, q_2, q_3)}_{A} \otimes \mbox{diag} (B^3). \qedhere
\end{eqnarray*}
\end{proof}
\vspace{-.25cm}
\begin{proof}[Proof of Thm~\ref{thm:ledges}]
\begin{eqnarray*}
\Delta_{C}^{(q_1, q_2, q_3)} & = & (\Pi_{C,q_2} C \Pi_{C,q_1}) \circ (C \Pi_{C,q_3} C) \\
& = &  \Big[ 
(\Pi_{A,q_2} \otimes I_B) (A \otimes B)(\Pi_{A,q_1} \otimes I_B) 
\Big] \\
&& \quad \circ \Big[
(A \otimes B)(\Pi_{A,q_3} \otimes I_B) (A \otimes B)
\Big] \\
& = & \Big[ 
(\Pi_{A,q_2}A \Pi_{A,q_1}) \otimes B 
\Big] \\
&& \quad \circ \Big[
(A \Pi_{A,q_3} A )\otimes (B^2)
\Big] \\
& = & 
\Delta^{(\uptau)}_{A} \otimes [B \circ (B^2)]. \qedhere
\end{eqnarray*}
\end{proof}

\bibliographystyle{IEEEtran}
\bibliography{aarabib}
%

\end{document}